\documentclass[english,preprint,pre]{revtex4}
\usepackage[T1]{fontenc}
\usepackage[latin9]{inputenc}
\usepackage{xcolor}
\usepackage{pdfcolmk}
\usepackage{amsmath}
\usepackage{graphicx}
\usepackage{amssymb}
\PassOptionsToPackage{normalem}{ulem}
\usepackage{ulem}

\makeatletter

\newcommand{\lyxmathsym}[1]{\ifmmode\begingroup\def\b@ld{bold}
  \text{\ifx\math@version\b@ld\bfseries\fi#1}\endgroup\else#1\fi}

\providecolor{lyxadded}{rgb}{0,0,1}
\providecolor{lyxdeleted}{rgb}{1,0,0}

\@ifundefined{textcolor}{}
{%
 \definecolor{BLACK}{gray}{0}
 \definecolor{WHITE}{gray}{1}
 \definecolor{RED}{rgb}{1,0,0}
 \definecolor{GREEN}{rgb}{0,1,0}
 \definecolor{BLUE}{rgb}{0,0,1}
 \definecolor{CYAN}{cmyk}{1,0,0,0}
 \definecolor{MAGENTA}{cmyk}{0,1,0,0}
 \definecolor{YELLOW}{cmyk}{0,0,1,0}
 }

\PassOptionsToPackage{caption=false}{subfig}
\newcommand{\vect}[1]{\boldsymbol{#1}}
\newcommand{\matr}[1]{\mathcal{#1}}

\usepackage{subfig}
\makeatother

\usepackage{babel}

\begin{document}

\title{Hydrodynamic Synchronisation of Model Microswimmers}

\author{V. B. Putz and J. M. Yeomans}

\affiliation{The Rudolf Peierls Centre for Theoretical Physics, Oxford University,
1 Keble Road, Oxford OX1 3NP, U.K.}

\date{\today}
\begin{abstract}
We define a model microswimmer with a variable cycle time, thus allowing
the possibility of phase locking driven by hydrodynamic interactions
between swimmers. We find that, for extensile or contractile swimmers,
phase locking does occur, with the relative phase of the two swimmers
being, in general, close to $0$ or $\pi$, depending on their relative
position and orientation. We show that, as expected on grounds of
symmetry, self T-dual swimmers, which are time-reversal covariant,
do not phase-lock. We also discuss the phase behaviour of a line of
tethered swimmers, or pumps. These show oscillations in their relative
phases reminiscent of the metachronal waves of cilia. 
\end{abstract}
\maketitle

\section{Introduction\label{sec:Introduction}}

The process of movement through a fluid poses particular challenges
to micron-scale swimmers, as their motion corresponds to the low Reynolds
number regime, where viscous forces dominate over inertia. A swimmer
moves by a cyclic deformation of its body and, at zero Reynolds number,
the resultant flow field is described by the Stokes equations which
have no time dependence. This kinematic reversibility of the flow
means that a swimmer's deformation must be non-reciprocal in time
to produce motion \cite{purcell_life_1977}. Non-reciprocal strokes
used by biological swimmers include rotating flagella, waving cilia
and surface deformations \cite{silverman_flagellar_1974,lighthill_flagellar_1976}.

A number of model swimmers have been devised which satisfy the requirement
for non-reciprocal deformations. These include swimmers made of joined
rods \cite{purcell_life_1977,becker_self-propulsion_2003}, spheres
with prescribed tangential velocities \cite{ishikawa_hydrodynamic_2006},
rigid dumbbells with phantom flagella \cite{hernandez-ortiz_transport_2005},
and linked spheres undergoing prescribed shape changes \cite{najafi_simple_2004,earl_modeling_2007,dreyfus_purcells_2005,avron_pushmepullyou:efficient_2005}
or subject to periodic forces \cite{keaveny_interactions_2008,gauger_numerical_2006,felderhof_swimming_2006}.
As a result of analytical and numerical work on the model systems
the swimming behaviour and efficiency of single swimmers is now rather
well characterised. Moreover, experiments approximating the simple
models are now becoming feasible using, for example, colloids manipulated
in optical traps \cite{leoni_basic_2008}. Much less is currently
understood about the way in which the hydrodynamic interactions between
two or more swimmers affect their motion. Pooley \emph{et al} \cite{pooley_swimming_2007}
have recently shown that the relative phase of two model swimmers
may be key in determining the way in which they interact. Numerical
work on large numbers of swimmers \cite{hernandez-ortiz_transport_2005,hernandez-ortiz_dynamics_2008,ishikawa_development_2008}
qualitatively reproduces the instabilities seen in experiments \cite{guell_hydrodynamic_1988,mendelson_organized_1999,dombrowski_self-concentration_2004,wu_particle_2000},
but links between the parameters describing individual swimmers and
the details of their collective behaviour are lacking.

The model swimmers which have been defined thus far in the literature
complete their stroke in a fixed time. This means that the relative
phase of two swimmers cannot be changed dynamically. However, given
the importance of relative phase in controlling swimmer-swimmer interactions,
it is feasible that biological swimmers could exploit phase changes
to control their relative motion. Indeed, there is evidence that hydrodynamic
phase synchronisation is feasible at low Reynolds number. Perhaps
the most striking example of this is the co-ordinated motion of beating
cilia, or metachronal waves, which is thought to be a consequence
of hydrodynamic interactions \cite{cosentino_lagomarsino_rowers_2002,lagomarsino_metachronal_2003,gueron_cilia_1997-1,gueron_energetic_1999}.
Moreover Taylor \cite{taylor_analysis_1951} demonstrated that two
undulating sheets minimise their dissipation if they oscillate in
phase and simulations have demonstrated that sperm cells adjust their
position to synchronise their motion \cite{yang_cooperation_2008}.
Reichert and Stark \cite{reichert_synchronization_2005} showed that
two rigid helices anchored in harmonic traps rotate in phase due to
hydrodynamic interactions.

Our aim in this paper is to demonstrate that low Reynolds number swimmers
can phase-lock as a result of the hydrodynamic interactions between
them. To do this we generalise one of the simplest model swimmers,
the linear three-sphere swimmer \cite{najafi_simple_2004}, to allow
a variable cycle period. We find that, for extensile or contractile
swimmers, phase locking does occur, with the relative phase of the
two swimmers being, in general, close to $0$ or $\pi$, depending
on their relative position and orientation. We show that, as expected
on grounds of symmetry, self T-dual swimmers, which are time-reversal
covariant, do not phase-lock. We also discuss the phase behaviour
of a line of tethered swimmers, or pumps. These show oscillations
in their relative phases reminiscent of the metachronal waves of cilia.

In Sec.~\ref{sec:Model} we define an extension to the linear three-sphere
swimmer which is designed to allow a variable cycle time. Sec.~\ref{sec:Equations-of-motion}
describes the corresponding equations of motion and the numerical
algorithm, based on the Rotne-Prager-Yamakawa approximation to zero
Reynolds number hydrodynamics, we use to solve them. Sec.~\ref{sec:Phase-locking-for}
demonstrates phase locking for two collinear swimmers, and the more
complex behaviour of three collinear swimmers is described in Sec.~\ref{sec:Three-collinear-swimmers}
and compared to that of tethered swimmers, or pumps, in Sec~\ref{sec:Tethered-collinear-swimmers}.
In Sec.~\ref{sec:Two-dimensions} we consider two swimmers moving
in a plane and show that they can lock-in to strokes with phase difference
$\sim0$ or $\pi$ depending on their relative positions.

\section{Model\label{sec:Model}}

\begin{figure}
\includegraphics[width=0.4\columnwidth]{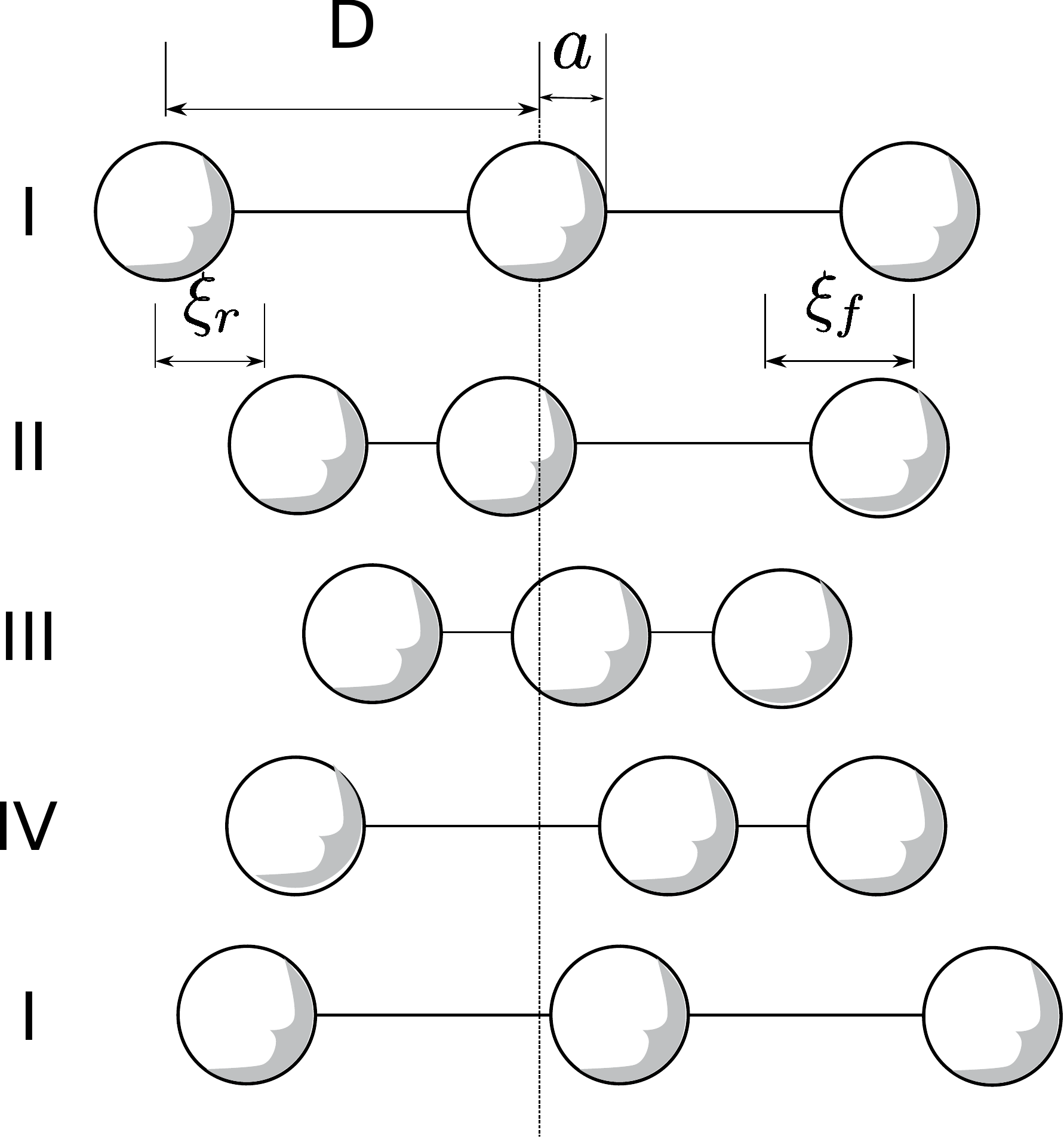} 

\caption{\label{fig:The-linear-three-sphere}Cycle of the linear three-sphere
swimmer; the nonreciprocal stroke permits motion at zero Reynolds
number.}

\end{figure}

We consider the linear, three-sphere swimmer illustrated in Fig.~\ref{fig:The-linear-three-sphere}
\cite{najafi_simple_2004}. The swimmer is made up of three spheres
of radius $a$ joined by thin rods of extended length $D$. The lengths
of the rods are altered to produce the swimming motion.

The swimming cycle proceeds in four distinct stages. First the left-hand
leg is retracted ($I\rightarrow II$ in Fig. \ref{fig:The-linear-three-sphere})
by a distance $\xi_{r}$. This results in hydrodynamic interactions
between the beads which move the swimmer a short distance to the left.
Then the right leg is retracted by $\xi_{f}$ ($II\rightarrow III$)
moving the swimmer to the right. The next two steps ($III\rightarrow IV\rightarrow I$)
extend the left-hand and right-hand leg in sequence, back to a length
$D$, and produce a rightwards, followed by a leftwards motion. Because
the steps which give motion to the right are performed with the passive
(constant length) leg contracted, the hydrodynamic interactions between
beads are stronger and hence the net motion is in this direction.

The motion of the spheres can be defined to take place at constant
velocity, and we shall term this a \emph{constant velocity swimmer}
\cite{golestanian_analytic_2007,pooley_swimming_2007,earl_modeling_2007}.
However it is also possible to drive the extension or contraction
of each rod by equal and opposite, constant forces $F_{a}$ acting
on the spheres at the end of that rod (a \emph{constant force swimmer}).
In this case, to ensure that within any given step of the cycle the
length of the passive leg is unchanged, equal and opposite constraint
forces must be defined to act at its ends.

Note that, for both the constant velocity and the constant force swimmers,
we have defined each stage of the cycle to continue until the active
rod has reached a specified \emph{length}, rather than lasting for
a specified \emph{time}. This is an irrelevant distinction for a single
swimmer because the relative position of the beads during the cycle,
rather than the speed with which it completes the stroke, defines
the swimmer motion; however, for more than one swimmer, the two definitions
are not equivalent because hydrodynamic interactions between swimmers
can affect how quickly a swimmer progresses through its swimming stroke.
Interacting constant velocity swimmers have, by definition, a fixed
cycle period. Interacting constant force swimmers, however, can reach
a specified leg length at different times, allowing the possibility
of phase synchronisation.

In general, the time-averaged far flow field around a swimmer is dipolar,
$\sim r^{-2}$, where $r$ is the distance from the swimmer. This
reflects the constraint that there are no external forces acting on
the swimmer; it is an active system, producing its own energy, For
the linear, three-sphere swimmer the long distance flow field is dipolar
as long as $\xi_{r}\neq\xi_{f}$. For $\xi_{r}<\xi_{f}$, the swimmer
is \emph{contractile} (a puller), pulling in fluid at its ends and
pumping it out at the sides. Conversely, for $\xi_{r}>\xi_{f}$, the
swimmer is \emph{extensile} (a pusher), and fluid is drawn in from
the sides and pushed out the ends. For the special case $\xi_{r}=\xi_{f}$,
the additional symmetry of the swimmer leads to a time-averaged flow
field which has quadrupolar symmetry and which decays as $r^{-3}$
at long distances \cite{pooley_swimming_2007}.

\section{Equations of motion\label{sec:Equations-of-motion}}

At sufficiently low Reynolds number that the inertial terms in the
Navier-Stokes equations can be neglected the fluid velocity \textbf{u}
obeys the Stokes equations \begin{equation}
\mu\nabla^{2}{\bf \vect{u}}-\nabla p=0\;,\label{eq:stokes}\end{equation}
 \begin{equation}
\nabla\cdot\vect{u}=0\;,\label{eq:continuity}\end{equation}
 where $\mu$ is the dynamic viscosity and $p$ is the pressure. Because
the Stokes equations are linear it is possible to write the velocity
field generated by an array of spheres as a linear combination of
the force ${\bf \vect{f}}_{m}$ acting on each sphere $m$ \cite{kirkwood_intrinsic_1948}.
This can be evaluated at the location of a sphere $n$ to give its
velocity \begin{equation}
{\bf \vect{\dot{q}}}_{n}=\sum_{m}\matr{H}_{nm}\vect{f}_{m}.\label{eq:Stokeslinear}\end{equation}
 The tensor $\matr{H}_{mn}$ is known exactly only for point spheres.
We use the Rotne-Prager-Yamakawa approximation \cite{rotne_variational_1969,yamakawa_transport_1970}
\begin{align}
\matr{H}_{mm}= & \left(1/6\pi\mu a\right)\matr{I},\label{eq:rpy_tensor_1}\\
\matr{H}_{mn}= & (1/8\pi\mu r_{mn})\left\{ \begin{array}{c}
\left[\matr{I}+\left(\vect{r}_{mn}\otimes\vect{r}_{mn}/r_{mn}^{2}\right)\right]\\
+\left(2a^{2}/r_{mn}^{2}\right)\left[\frac{1}{3}\matr{I}-\left(\vect{r}_{mn}\otimes\vect{r}_{mn}/r_{mn}^{2}\right)\right]\end{array}\right\} \label{eq:rpy_tensor}\end{align}
 where $\vect{r}_{mn}$ is the vector between the two spheres, $m$
and $n$. This is expected to be a good approximation for $r_{mn}\gtrsim2a$
\cite{yamakawa_transport_1970}. 

In general Eqn.~(\ref{eq:Stokeslinear}) must be solved numerically.
The inputs are the forces acting on the spheres bounding the active
leg together with the constraint that the passive leg does not change
its length. The outputs are the velocities of the beads, which in
turn define the swimming motion. Imposing a fixed length on the passive
leg is equivalent to imposing constraint forces on the beads at its
ends. These must be equal and opposite as they are forces internal
to the swimmer. In two or more dimensions, additional constraint forces
are also needed to preserve the linearity of the swimmers.

\begin{figure}
\begin{centering}
\includegraphics[width=0.4\columnwidth]{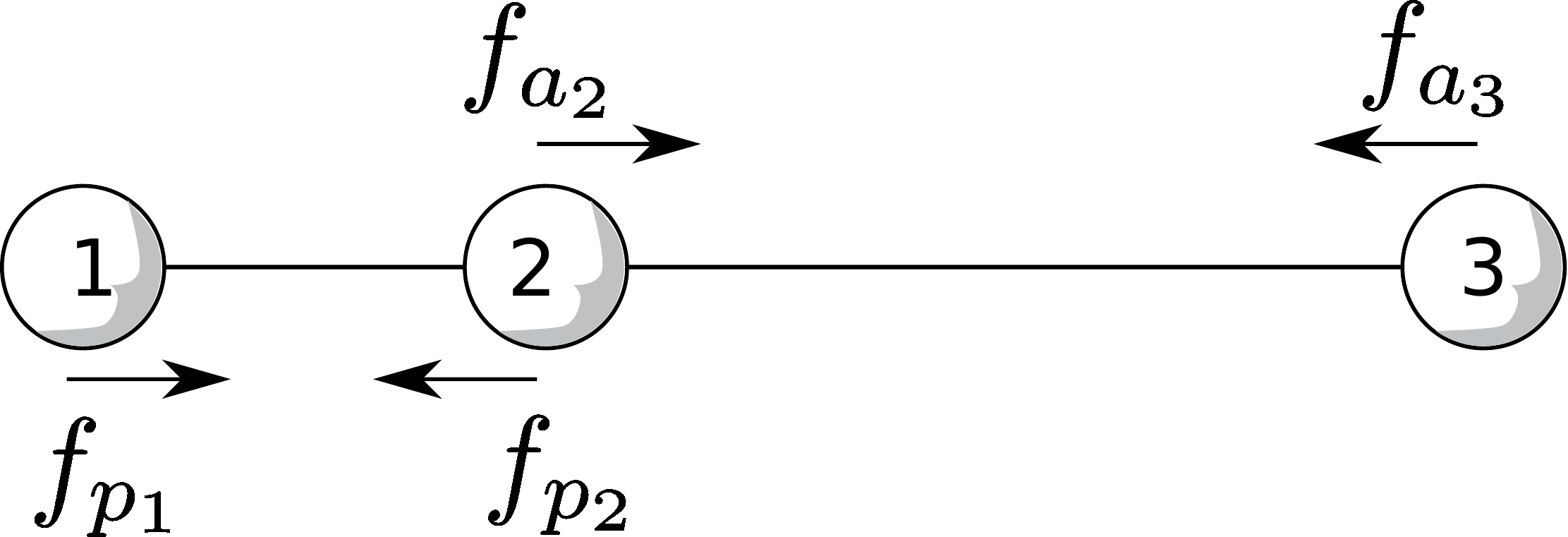}
\par\end{centering}

\centering{}\caption{Forces on a three-sphere swimmer during contraction of the right-hand
leg.\label{fig:Constraint-forces-on}}
 
\end{figure}

To obtain an expression for the constraint forces, we define the positions
of the $3N$ spheres needed to construct $N$ swimmers by the $9N$-component
vector $\vect{q}$. We define $\vect{f}_{a}\equiv\left(\vect{f}_{a1},\vect{f}_{a2}\ldots\right)$
as the $9N$-component vector listing the (known) forces moving the
active legs of the swimmers and vectors $\vect{f}_{p\alpha}\equiv\left(\vect{f}_{p1},\vect{f}_{p2}\ldots\right)$
describing the forces needed to impose each of the constraints which
preserve the lengths of the passive legs. For example, for swimmer
one, when the leg between beads 1 and 2 remains at a constant length,
$\vect{f}_{p\alpha}=\left(\vect{f}_{p1},\vect{f}_{p2},0,0,0\ldots\right)$
(see Fig.~\ref{fig:Constraint-forces-on}). Each constraint $\alpha$
is specified by a function $C_{\alpha}(\vect{q})=0$. $\dot{C}_{\alpha}=\sum_{i}\frac{\partial C_{\alpha}}{\partial q_{i}}\dot{q}_{i}=0$,
or, defining $j_{\alpha i}\equiv\frac{\partial C_{\alpha}}{\partial q_{i}}$,\begin{equation}
\vect{j}_{\alpha}^{T}\cdot\dot{\vect{q}}=0.\label{eq:j_dot_q_equals_0}\end{equation}

The constraint forces must do no work. Therefore\begin{equation}
\vect{f}_{p\alpha}^{T}\cdot\dot{\vect{q}}=0.\label{eq:contstraint_does_no_work}\end{equation}
Eqns.~(\ref{eq:j_dot_q_equals_0}) and (\ref{eq:contstraint_does_no_work})
imply that\begin{equation}
\vect{f}_{p\alpha}=\lambda_{\alpha}\vect{j}_{\alpha}.\label{eq:forces_are_multiples_of_jac}\end{equation}
Substituting Eqn.~(\ref{eq:forces_are_multiples_of_jac}) into Eqn.~(\ref{eq:Stokeslinear}),
and using Eqn.~(\ref{eq:j_dot_q_equals_0}) gives\begin{equation}
-\vect{j}_{\beta}^{T}\matr{\mathcal{H}}\vect{f}_{a}=\sum_{\alpha}\vect{j}_{\beta}^{T}\matr{\mathcal{H}}\vect{j}_{\beta}\lambda_{\alpha},\label{eq:jhf=jhjtlambda}\end{equation}
a set of simultaneous equations that can be solved for $\lambda_{a}.$
In this case $\vect{j}_{\beta}^{T}\matr{\mathcal{H}}\vect{j}_{\alpha}$
is symmetric, and hence\begin{equation}
\lambda_{\alpha}=-\left(\vect{j}_{\alpha}^{T}\matr{\mathcal{H}}\vect{j}_{\beta}\right)^{-1}\left(\vect{j}_{\alpha}^{T}\matr{\mathcal{H}}\vect{f}_{a}\right).\label{eq:lambda=jhjt-1*jhf}\end{equation}

\section{Phase locking of one-dimensional swimmers\label{sec:Phase-locking-for}}

The motion of each swimmer is controlled by fixed applied forces $\vect{f}{}_{a}$
which act until the active arm reaches a prescribed length. When more
than one swimmer is present external advection produced by fluid flow
from other swimmers' motions can have an impact on how \emph{quickly}
the swimmer proceeds through its cycle. This allows for the possibility
of phase-locking.

To discuss the notion of phase in a non-sinusoidal context, and for
a variable-length cycle, it is most convenient to define a \emph{spatial
phase} which is a function only of the lengths of a swimmer's legs.
During a complete swimming stroke, the active spheres move a total
distance $2(\xi_{f}+\xi_{r})$ relative to the centre sphere. We define
the spatial phase $\phi$ of a swimmer as the fraction of this distance
travelled since the beginning of the stroke multiplied by 2$\pi$,
giving a number varying from $0$ at the beginning of a stroke to
$2\pi$ at the completion of a stroke.

As a first example we consider the behaviour of two collinear swimmers,
which will move in one dimension. Two extensile swimmers, with $\xi_{f}=0.1$
and $\xi_{r}=0.5$, were placed along the $x$-axis with their centres
of mass separated by $3D$. The initial phase of the leading swimmer
was taken to be $\phi=0$ and several simulations were run for different
initial phases of the trailing swimmer. The simulations were run for
50,000 swimmer cycles.

\begin{figure}
\subfloat[]{\includegraphics[width=0.4\columnwidth]{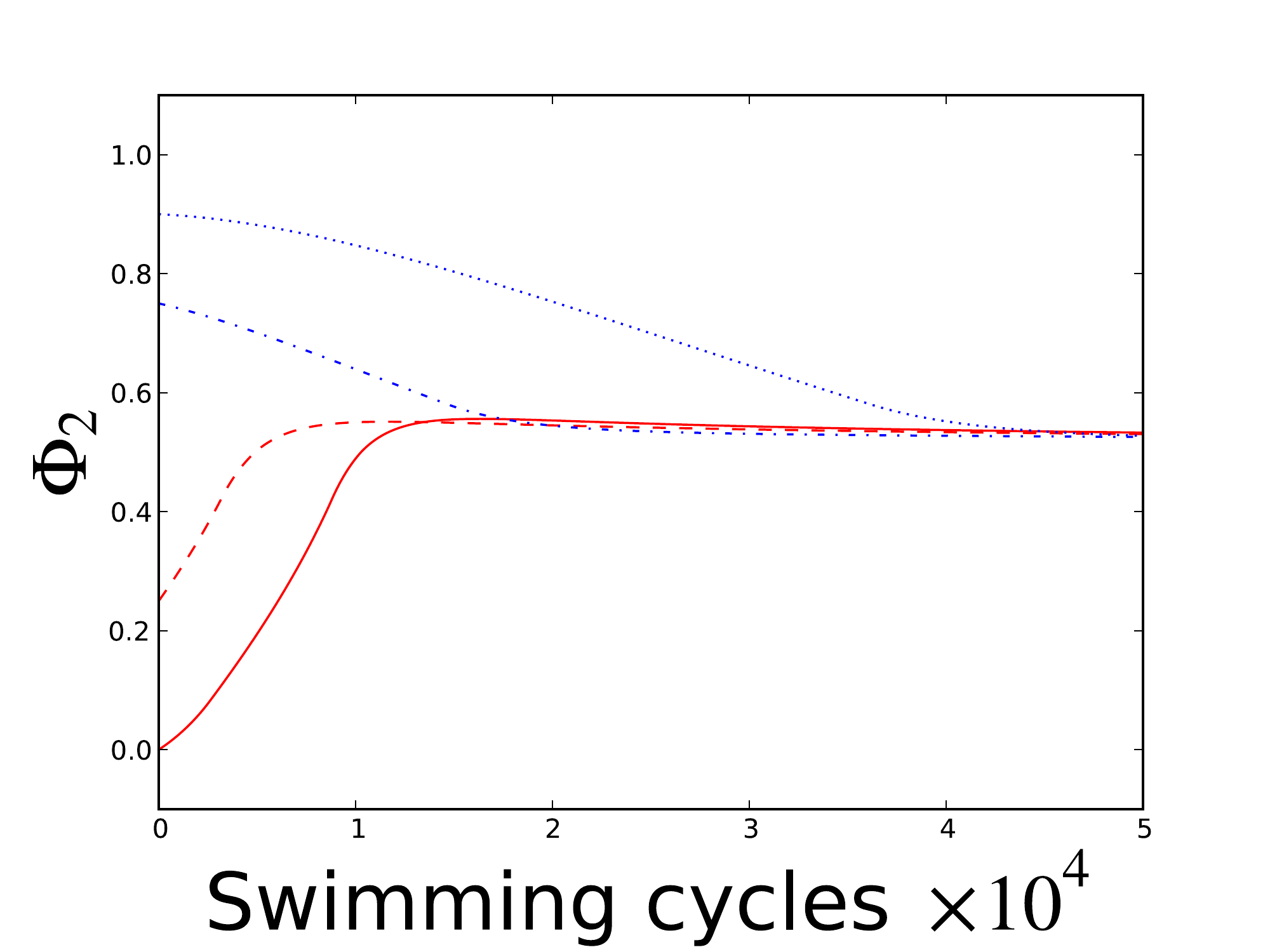}} \subfloat[]{\includegraphics[width=0.4\columnwidth]{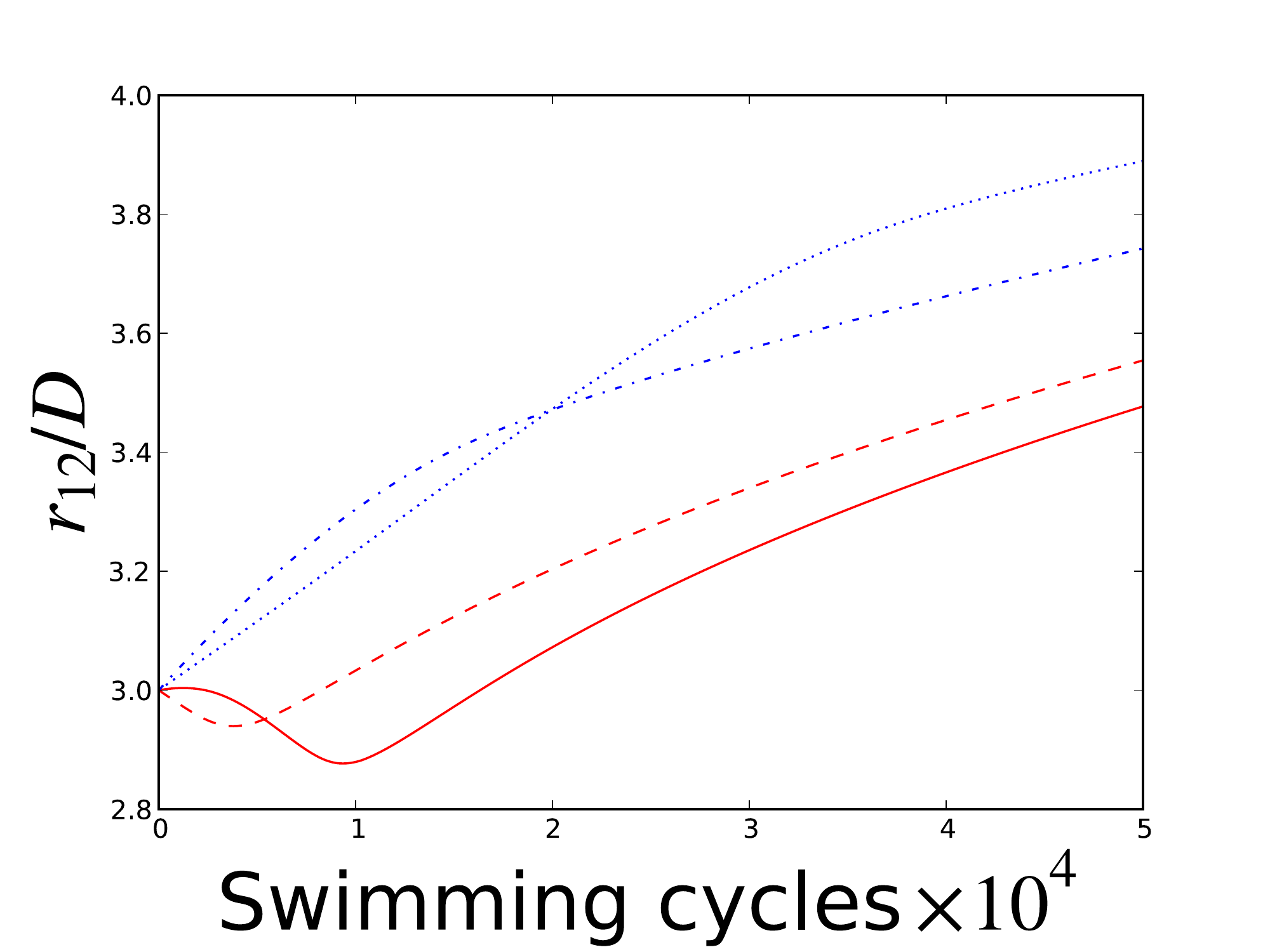}}

\caption{\label{fig:Phase-locking-behaviour}Time evolution of the motion
of two collinear, extensile swimmers with $D=1,a=0.03,\xi_{r}=0.5,\xi_{f}=0.1$
for different initial relative phase. (a) Variation of the relative
phase with time: the graph shows the phase of the rear swimmer, $\phi_{2}$,
measured at the beginning of the stroke of the leading swimmer ($\phi_{1}=0$).
(b) Distance between the centres of mass of the swimmers.}

\end{figure}

Fig.~\ref{fig:Phase-locking-behaviour}(a) compares the variation
of the phase of the rear swimmer, measured at the beginning of the
lead swimmer's stroke, with time. It is apparent that the phase difference
between the two swimmers slowly drifts toward a stable value, indicating
phase-locking behaviour. At the lock-in point the swimmers are approximately
out of phase with each other ($\phi\approx\pi$).

As the relative phase changes the transport of the swimmer is affected.
The evolution of the distance between the two swimmers is plotted
in Fig. \ref{fig:Phase-locking-behaviour}(b) . As the trailing swimmer
changes its phase it moves closer to the lead swimmer for a time before
settling, at the locking point, to motion away from the lead swimmer.
This is in agreement with the conclusion in \cite{pooley_swimming_2007}
that extensile, collinear, constant period swimmers attract if they
are in phase but repel if they are out of phase.

\begin{figure}
\subfloat[]{\includegraphics[width=0.4\columnwidth]{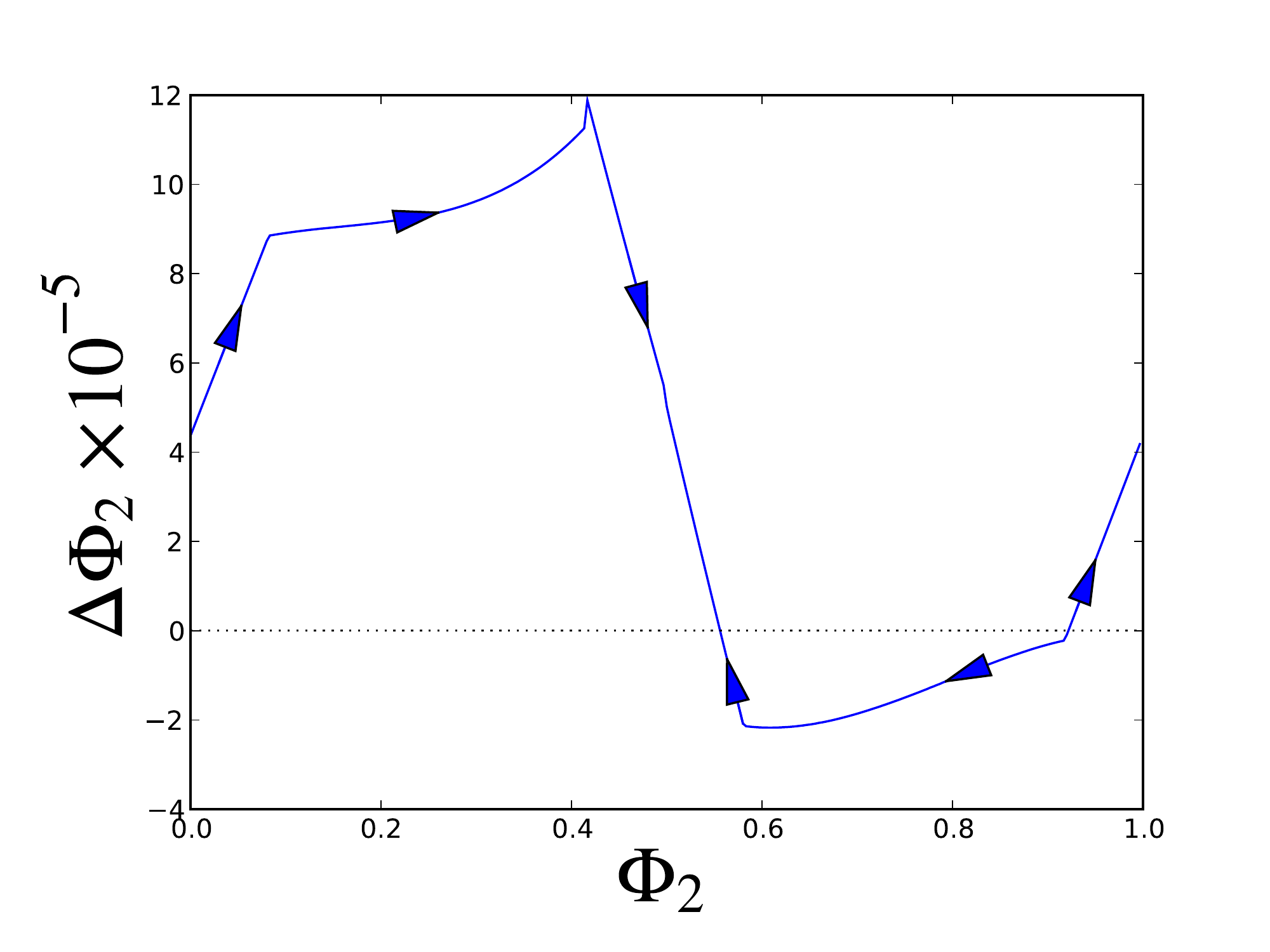}} \subfloat[]{\includegraphics[width=0.4\columnwidth]{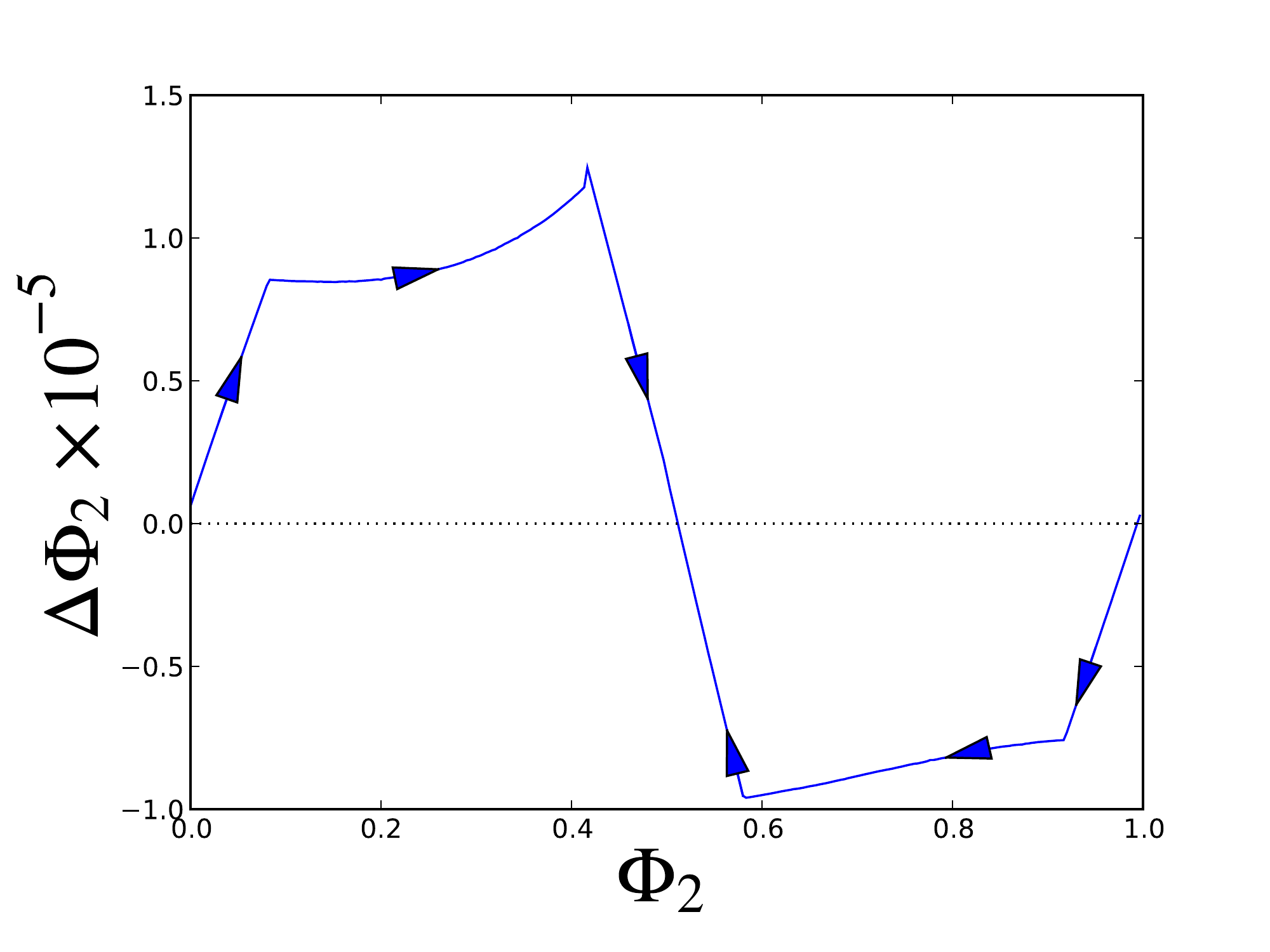}}

\subfloat[]{\includegraphics[width=0.4\columnwidth]{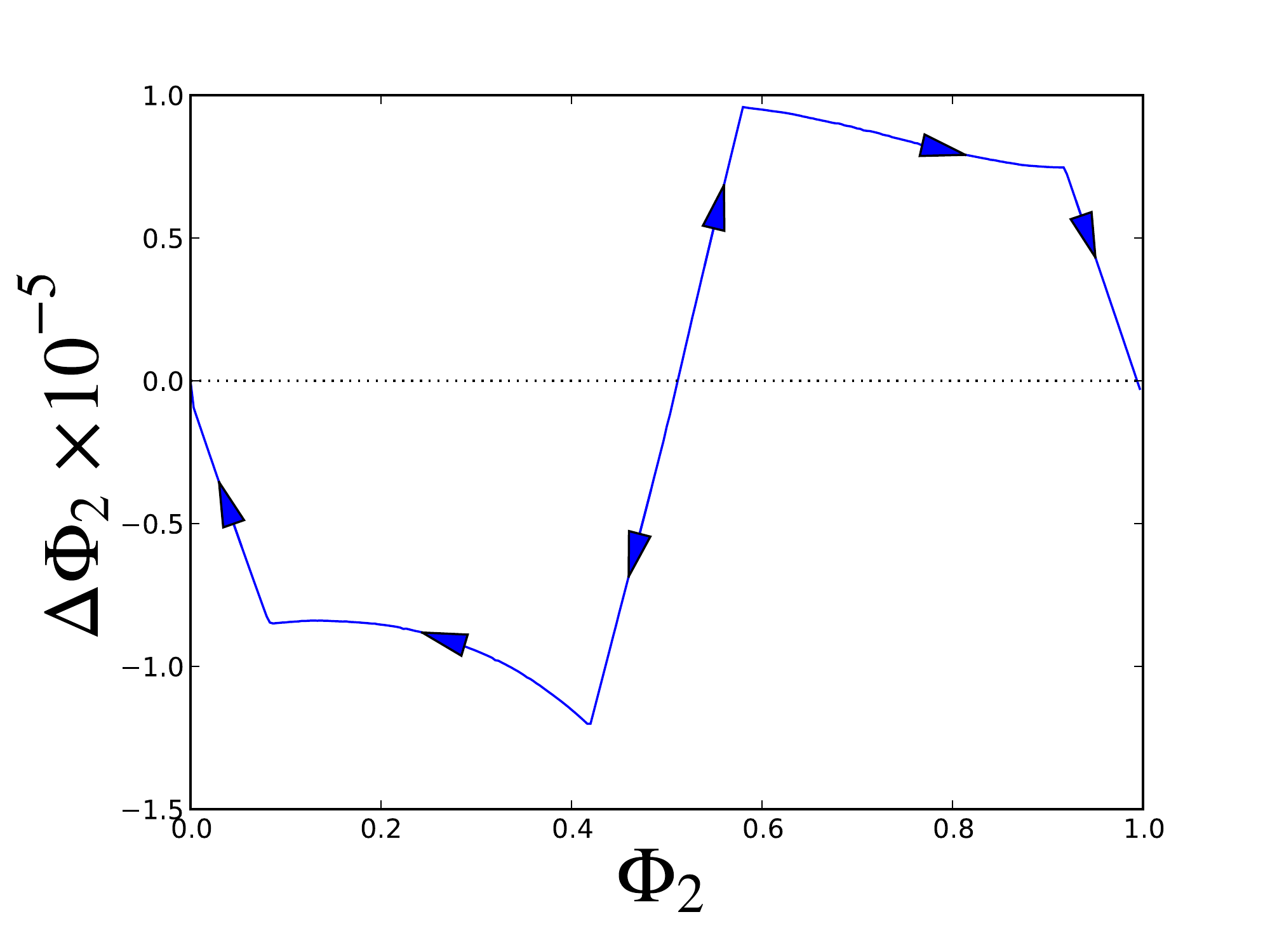}} \subfloat[]{\includegraphics[width=0.4\columnwidth]{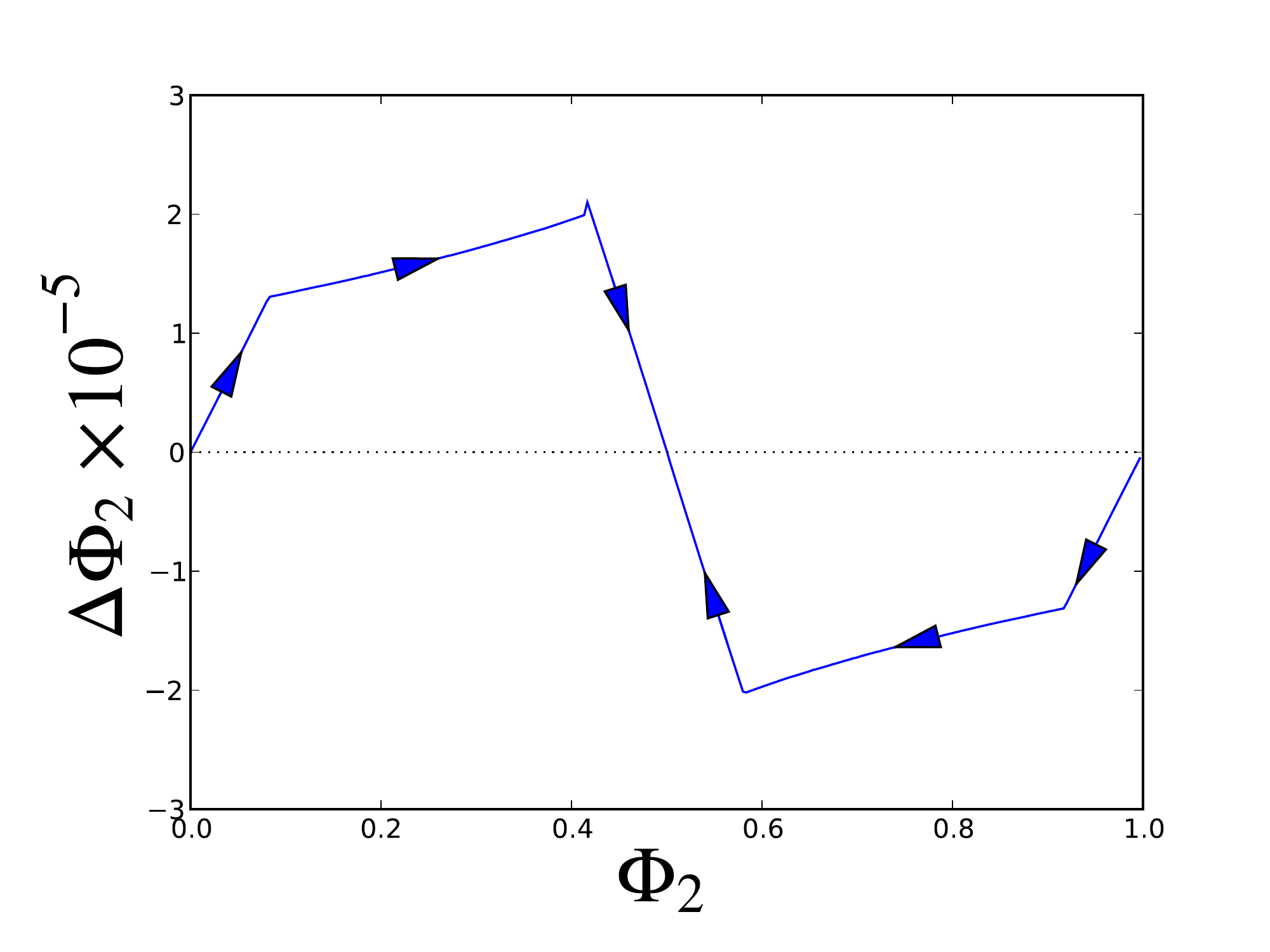}}

\caption{\label{fig:Comparing-relative-cycle}Change in the phase of the trailing
swimmer after one swimming cycle as a function of its phase at the
beginning of the cycle of swimmer 1. Crossing points on the $x$-axis
represent either an unstable (if the slope is positive) or a stable
lock-in point (if the slope is negative). Arrows indicate how the
swimmer's motion evolves with time. All swimmers have $\xi_{r}=0.5,$
$\xi_{f}=0.1,\lyxmathsym{ }D=1$. (a) separation$=3D$; extensile,
$\xi_{r}=0.5,$~$\xi_{f}=0.1$; swimmers moving in same direction;
(b) separation$=5D$; extensile, $\xi_{r}=0.5,$ $\xi_{f}=0.1$; swimmers
moving in same direction; (c) separation=$5D$; contractile, $\xi_{r}=0.1$,
$\xi_{f}=0.5$; swimmers moving in same direction; (d) separation=$5D$;
extensile, $\xi_{r}=0.5,$ $\xi_{f}=0.1$; swimmers moving away from
each other.}

\end{figure}

The phase locking occurs because velocity gradients across a swimmer
due to hydrodynamic interactions between swimmers can either assist
or hinder the leg motion. Fig.~\ref{fig:Comparing-relative-cycle}
illustrates this by showing the change in phase experienced by the
trail swimmer after one cycle. If the trail swimmer is iterating through
its cycle more quickly than the lead swimmer, it will show a positive
change in phase; this corresponds to moving along the graph in Fig.
\ref{fig:Comparing-relative-cycle} to the right. Similarly, if the
trail swimmer is hindered in its cycle, its phase will fall behind
that of the lead swimmer, corresponding to moving to the left. Locking
will occur when a stable point (zero phase shift) is reached. The
further the curve from the $x$-axis, the faster the phase-locking
behaviour proceeds. Details of the shape of the phase-change curve
in Fig.~\ref{fig:Comparing-relative-cycle} depend on the parameters
of the swimmers and their separation. In general it becomes flatter,
and more centred about a phase change of zero, as the distance between
the swimmers increases, because the hydrodynamic interactions between
them decrease. 

\begin{figure}
\includegraphics[width=0.4\columnwidth]{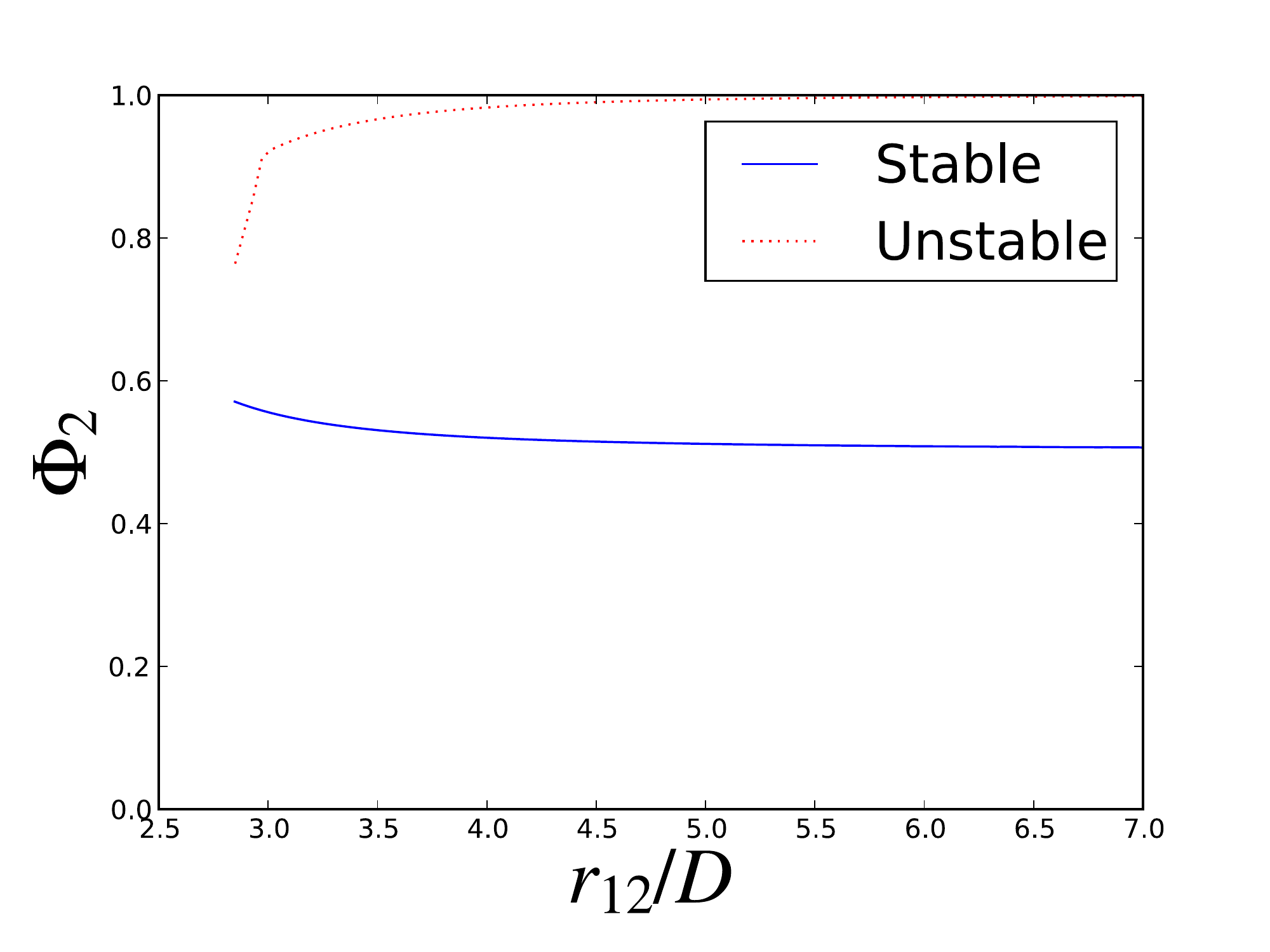}

\raggedright{}\caption{\label{fig:Stable-and-unstable}Stable and unstable lock-in points
of the relative phase of two collinear swimmers as a function of their
separation. }

\end{figure}

By examining the points at which the curve crosses the $x$-axis for
increasing swimmer-swimmer spacing, we can identify how the relative
phase of the swimmers in the phase-locked state varies with their
separation. This is shown in Fig.~\ref{fig:Stable-and-unstable}.
For a collinear arrangement of two swimmers, the lock point depends
only weakly on the distance between them, although the time it takes
to adjust phases increases with separation. For the case of collinear
swimmers moving in opposite directions (Fig.\ref{fig:Comparing-relative-cycle}(d))
the symmetry of the arrangement forces the stable lock point to be
exactly halfway through the cycle ($\phi_{2}=\pi$).

The swimmer we have considered so far has leg amplitudes $\xi_{f}<\xi_{r}$.
Such an extensile swimmer is related to a contractile swimmer, with
$\xi_{f}$ and $\xi_{r}$ interchanged, through a T-dual transformation
ie time reversal, together with a relabelling of front and rear \cite{alexander_scattering_2008}.
In terms of Fig.~\ref{fig:Comparing-relative-cycle} this corresponds
to an inversion of the curve $y\rightarrow-y$ (compare Figs. \ref{fig:Comparing-relative-cycle}(b)
and (c)). Hence the stable zeros of an extensile swimmer become the
unstable zeros of its T-dual, contractile counterpart, and contractile
swimmers lock-in to a relative phase $\phi\sim0$.

When $\xi_{r}=\xi_{f}$, the swimmer is neither extensile nor contractile
and it maps onto itself under the T-duality transformation (its flow
field is not dipolar, but decays more rapidly, as $r^{-3}$ \cite{pooley_swimming_2007}).
The phase-change curve now lies along the $x$ axis, corresponding
to a difference in cycle time of zero for all relative phases. Hence
there is no phase-locking, as expected on grounds of symmetry, for
a swimmer which is self T-dual.

\section{Three collinear swimmers\label{sec:Three-collinear-swimmers}}

\begin{figure}
\subfloat[]{\includegraphics[width=0.8\columnwidth]{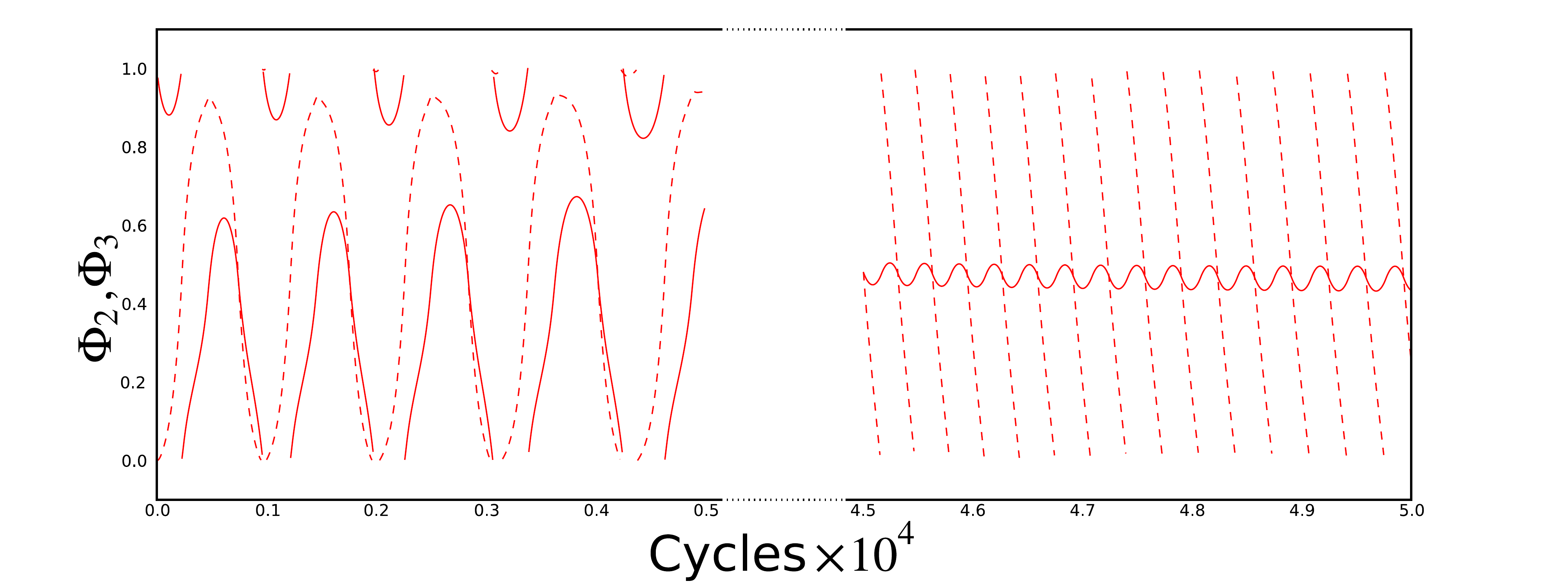}}

\subfloat[]{\includegraphics[width=0.4\columnwidth]{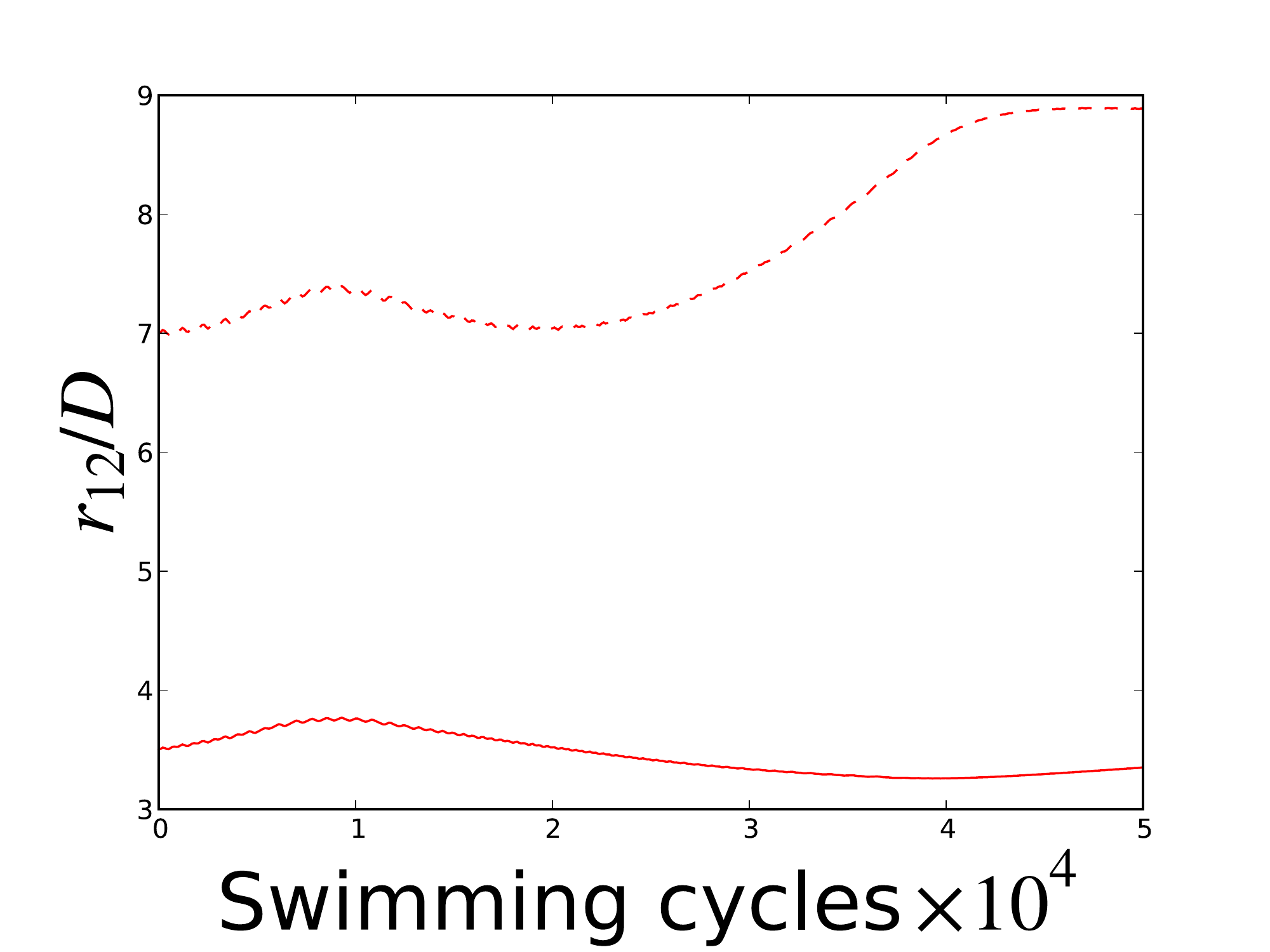} }\subfloat[]{\includegraphics[width=0.4\columnwidth]{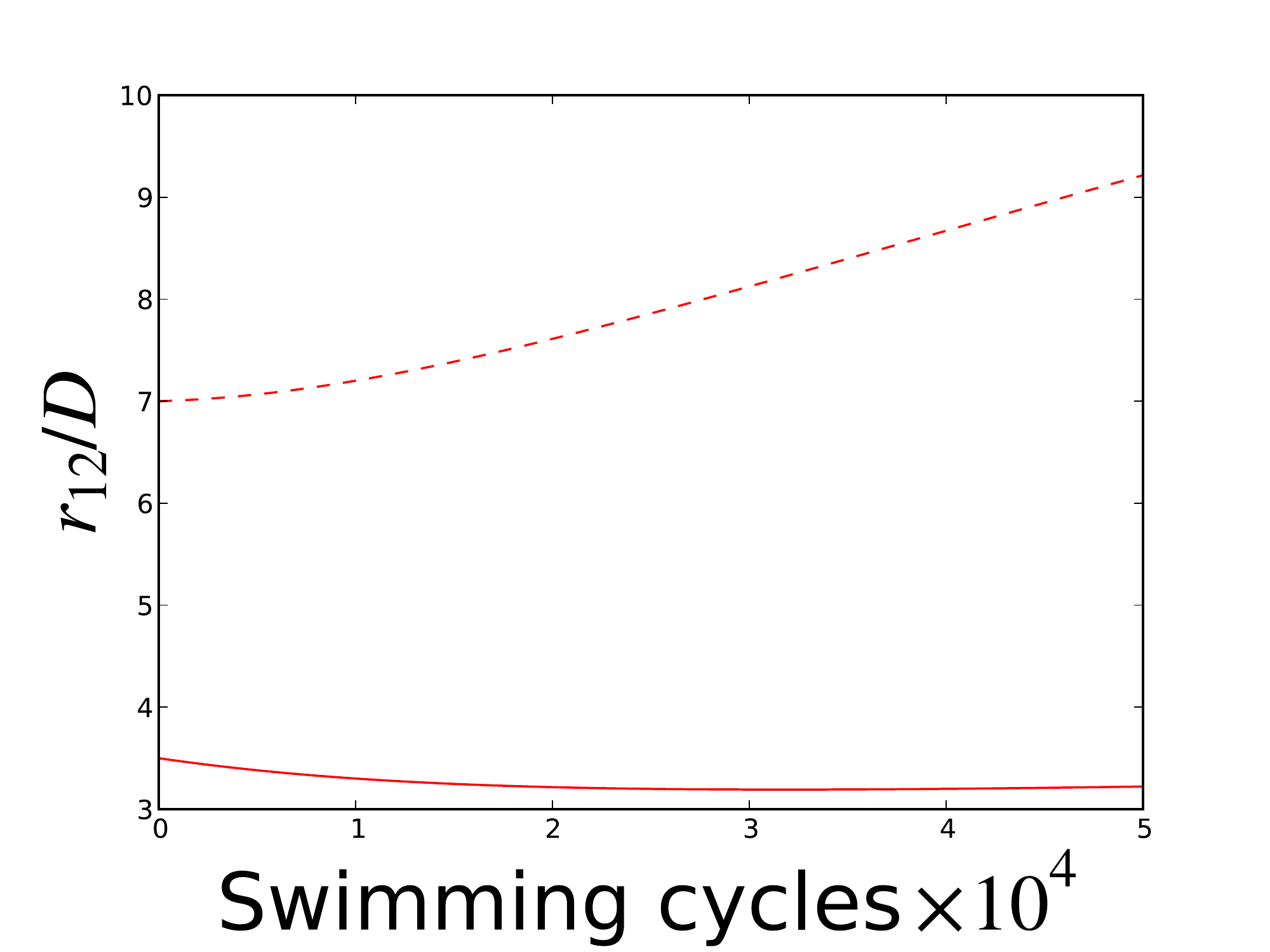}
}

\caption{Time evolution of the relative phase of three collinear swimmers\label{fig:Chaotic-behaviour-of}.
(a) relative phase of the middle and rear swimmers at the beginning
of the front swimmer's cycle (b) displacements of the middle and rear
swimmers from the front swimmer in a constant force swimmer simulation
(c) displacements of the middle and rear swimmers from the front swimmer
in a constant velocity swimmer simulation.}

\end{figure}

We have demonstrated that hydrodynamic phase locking is possible for
two collinear swimmers. We now investigate the behaviour of three
swimmers, again moving in one dimension. Consider three collinear,
extensile swimmers, initially with spacing 2.5D, and all initially
with $\phi=0$. Fig.~\ref{fig:Chaotic-behaviour-of} shows the phases
of the middle and rear swimmer when the lead swimmer reaches the beginning
of its cycle as a function of the number of swimmer cycles. It is
apparent from this figure that there is no simple phase locking. Early
in the simulation, the phases of swimmers 2 and 3 relative to that
of swimmer 1 do not tend to a fixed point as for the two-swimmer case,
but vary periodically.

As the simulation proceeds, the hydrodynamic interaction between the
swimmers acts to change their relative positions. Swimmers 1 and 2
remain at a similar spacing, and swimmer 3 drifts away as shown in
Fig.~\ref{fig:Chaotic-behaviour-of}(b). The stronger interaction
between swimmers 1 and 2 constrains their relative phase to be close
to $\pi,$ but with a small oscillation imposed by the third swimmer,
as its phase decreases at an approximately constant rate.

For comparison we present, in Fig~\ref{fig:Chaotic-behaviour-of}(c),
similar results for the relative positions of constant velocity swimmers
which remain in phase. There is the same tendency for one of the swimmers
to drift away from the other two. At the longest time this swimmer
is still moving away, whereas for the constant force case, its position
has been stabilized by the changing relative phase.

\section{\label{sec:Tethered-collinear-swimmers}Tethered collinear swimmers}

Swimmers which are constrained to remain in a fixed position become
pumps that drive a net flow field. The flow field and hydrodynamic
interactions will be changed by the constraint; in particular the
far flow field will, in general, decay as $r^{-1}$ because of the
force holding the swimmer in position. The locking is likely to be
simpler, as the distance between the pumps, and hence the strength
of the hydrodynamic interactions between them, will remain constant.

To investigate this we considered collinear swimmers, whose centre
of mass was fixed in space by applying a small external force to each
sphere at each time step of the simulation. The force necessary to
do this was relatively small, $\sim10^{-3}$ of the magnitude of the
internal forces necessary to move the legs. 

\begin{figure}
\subfloat[]{\includegraphics[width=0.8\columnwidth]{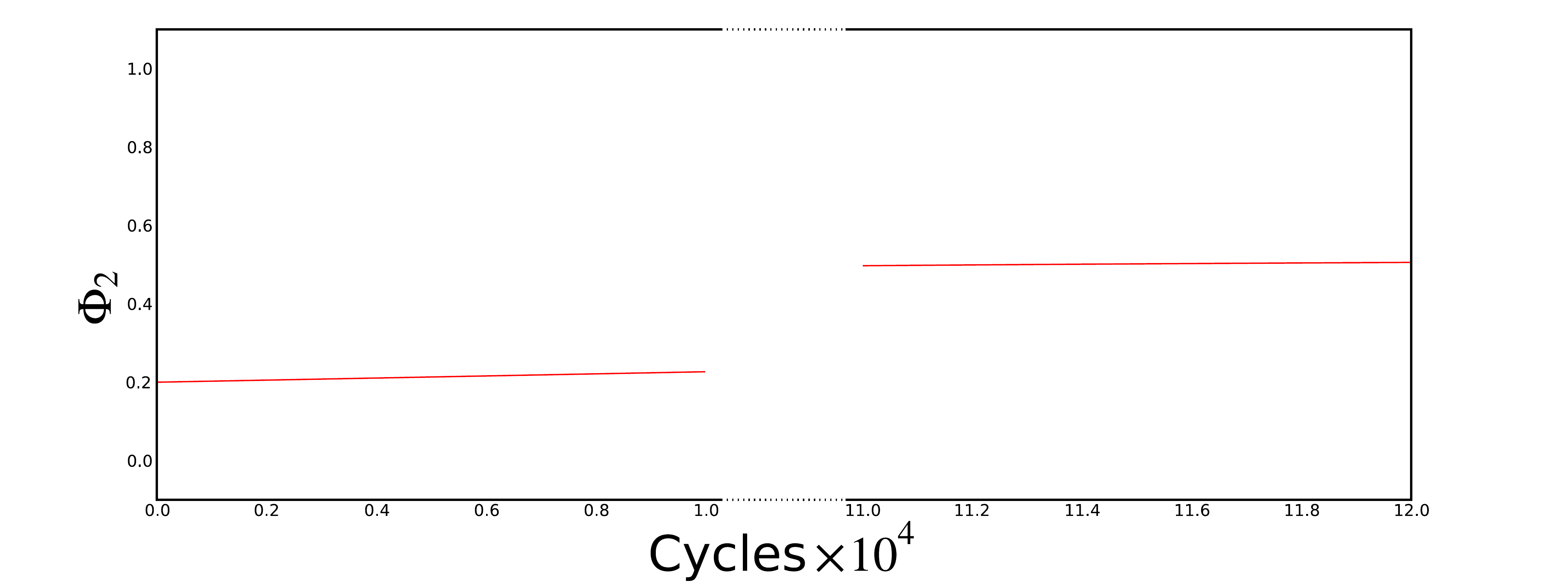}}

\subfloat[]{\includegraphics[width=0.8\columnwidth]{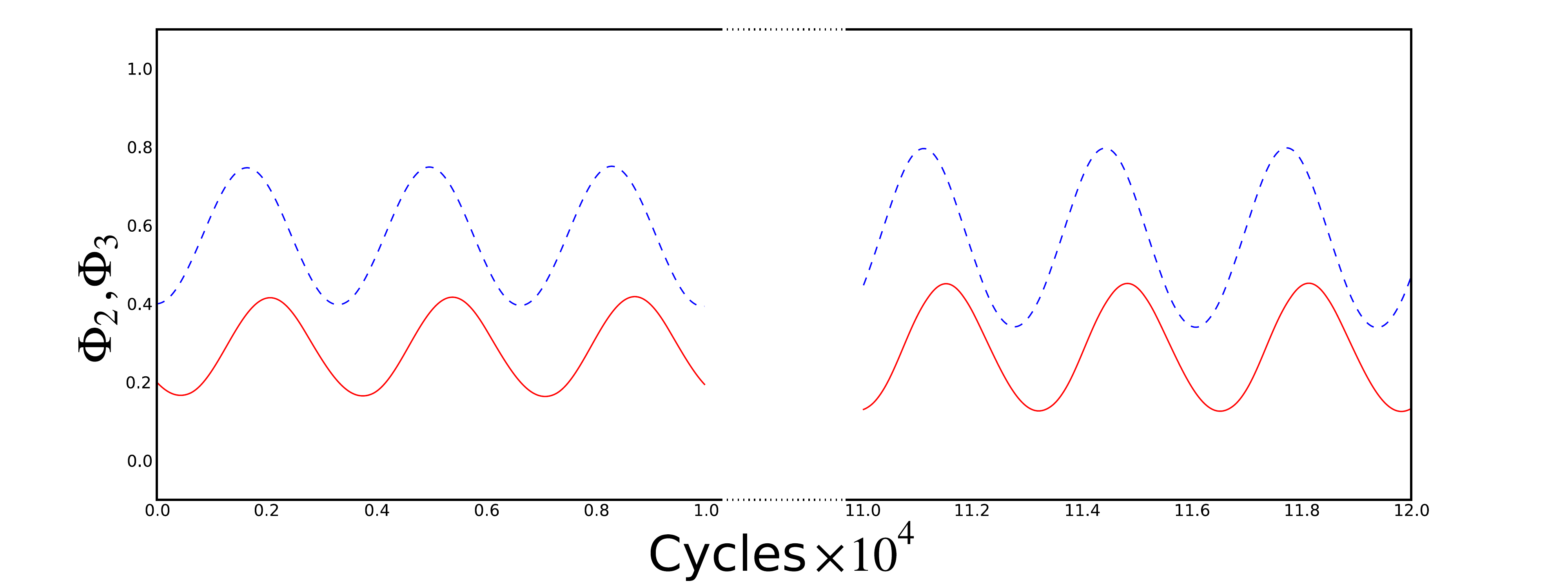}

}

\subfloat[]{\includegraphics[width=0.8\columnwidth]{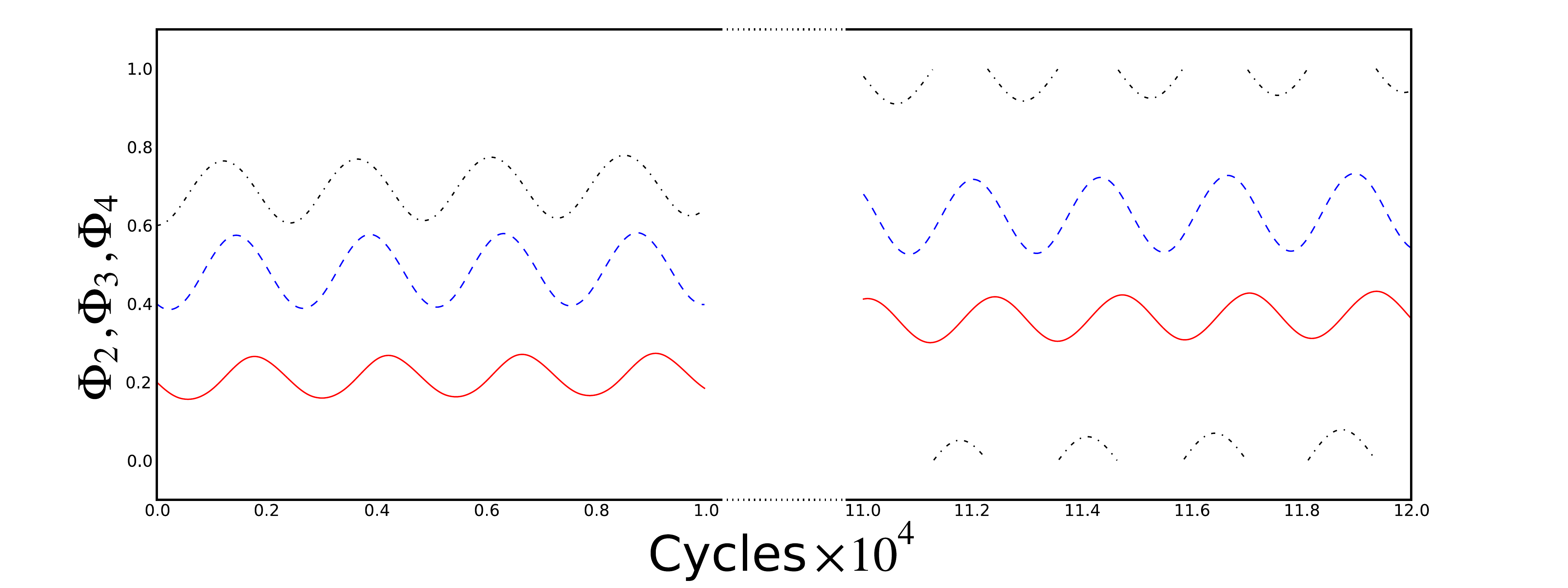}}

\caption{\label{fig:Multiple-pinned-swimmers}Time evolution of the relative
phases of collinear pumps with time (a) two (b) three (c) four swimmers.
The phases are measured at the beginning of the cycle of pump 1}

\end{figure}

The evolution of the relative phases of the pumps with time is shown
in Fig. \ref{fig:Multiple-pinned-swimmers}. Two extensile pumps phase-lock
with a phase difference of $\pi$, as for the free swimmers. Three
or four equispaced pumps show stable oscillations in their relative
phases. These are reminiscent of the metachronal waves observed in
arrays of cilia \cite{cosentino_lagomarsino_rowers_2002,lagomarsino_metachronal_2003,gueron_cilia_1997-1,gueron_energetic_1999}.

\section{Two dimensions\label{sec:Two-dimensions}}

\begin{figure}
\includegraphics[width=1\columnwidth]{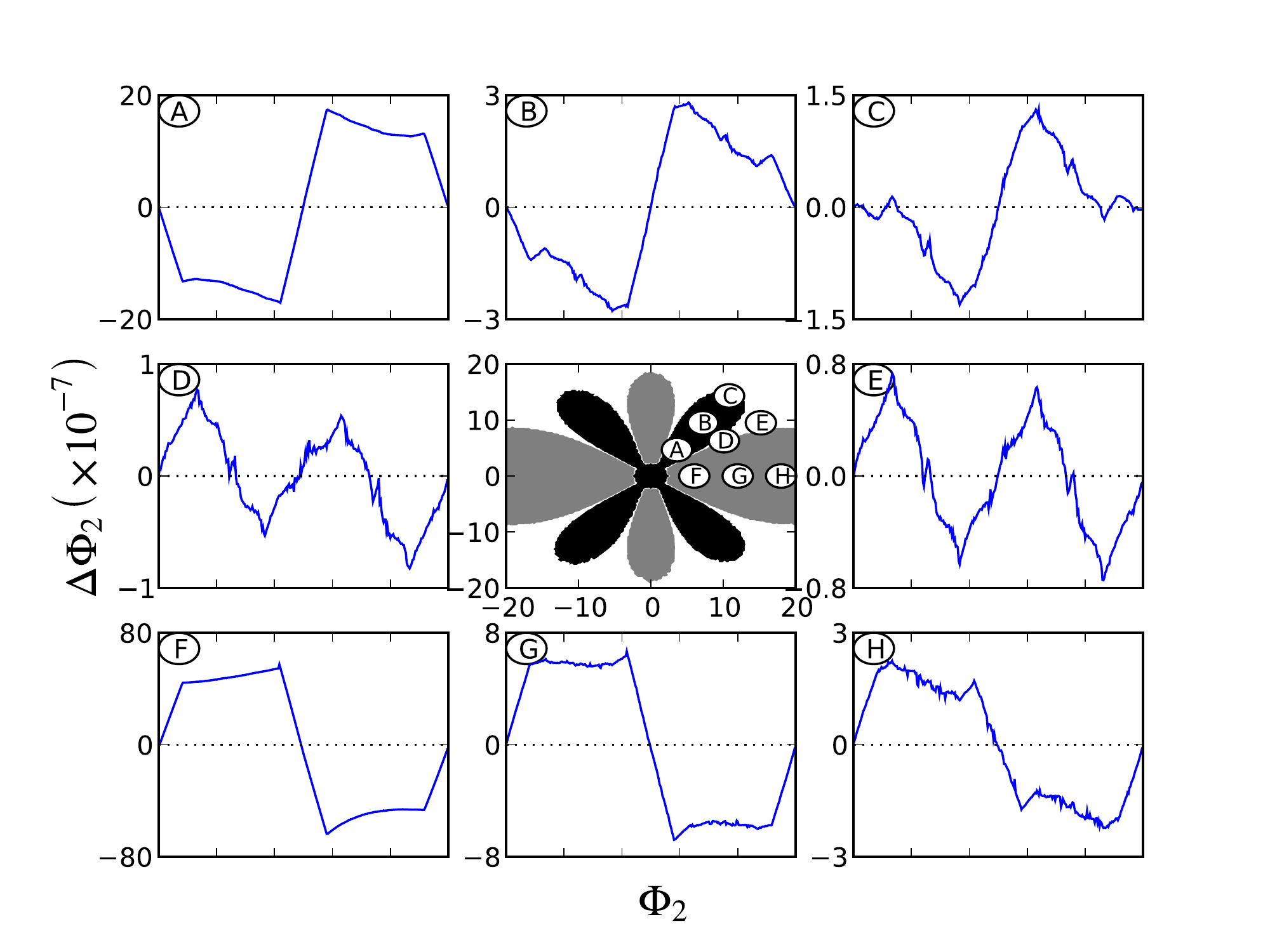}

\caption{Centre: phase locking behaviour of two extensile swimmers. The swimmers
are parallel and swimmer 1 is at the origin. The axes are labelled
in units of $D$. When swimmer 2 lies in the grey areas, the swimmers
synchronise out-of-phase ($\phi=\pi$). When swimmer 2 lies in the
black areas, the swimmers synchronise in-phase ($\phi=0$). The white
areas represent regions with two or more stable zeros. Sides: phase
change curves corresponding to swimmer configurations $A,B\ldots H$.
In each case the change of phase of swimmer 2 per cycle is plotted
as a function of the phase of swimmer 2 measured at the beginning
of the stroke of swimmer 1.\label{fig:Grid-image-showing}}

\end{figure}

The phase-locking behaviour of two swimmers is considerably more complex
when they are not collinear. To investigate this we ran simulations
with two extensile swimmers, initially parallel, for different distances
apart and for different angles $\theta$ between the swimmer orientation
and the vector joining their centres. For each position, we ran the
swimmers through one cycle at relative phases from $0$ to $2\pi$
to produce phase-change curves analogous to those in Fig. \ref{fig:Comparing-relative-cycle}.
Examples are shown in Fig.~\ref{fig:Grid-image-showing}, together
with a central panel summarising the locking behaviour as a function
of swimmer spacing and $\theta$.

As $\theta$ varies, the phase change curves change dramatically.
Consider first increasing $\theta$ along the arc $GDB$ in Fig. \ref{fig:Grid-image-showing}.
At $G$ ($\theta=0$), there is lock-in to a phase difference $\phi=\pi$,
corresponding to the the swimmers oscillating in antiphase. As $\theta$
increases, the fixed point loses its stability via a bifurcation and
two new stable fixed points appear ($D$). These move apart and annihilate,
such that stability is transferred to the fixed point at $\phi=0$,
and the swimmers synchronize in phase ($B$).

This interchange of stability occurs several times as $\theta$ varies,
giving regions of lock-in which are alternately in-phase and out-of-phase.
The pattern reflects the dipolar symmetry of the hydrodynamic interactions.

As the swimmers move further apart, remaining parallel but at the
same angle $\theta,$ (e.g. $FGH$ in Fig. \ref{fig:Grid-image-showing}),
the curves retain approximately the same shape but the change in phase
per cycle is smaller, reflecting the weakening interactions. Beyond
the shaded regions two or more lock points are seen but it is difficult
to distinguish locking from numerical noise.

\section{Discussion\label{sec:Discussion}}

We have extended the definition of a simple linear, three-sphere swimmer
model to permit variable stroke periods, and hence to allow for the
possibility of phase synchronisation at zero Reynolds number. We find
phase locking for two swimmers, and more general cooperative phase
behaviour for more than two swimmers. The phase-locking is driven
by hydrodynamic interactions.

In general the swimmers lock-in to a phase difference of $\sim0$
or $\sim\pi$, depending on their relative positions, although more
complicated synchronisation behaviour is possible. For three swimmers
the relative phases oscillate, together with a superimposed drift
in time, as the swimmer positions vary. The locking is slow, tens
or hundreds of swimmer cycles, becoming slower with increasing separation.

A similar synchronisation is seen for tethered swimmers (pumps), although
the behaviour is simpler because the relative positions of the pumps
do not vary. Two collinear pumps lock-in to a phase difference $0$
or $\pi$ for contractile or extensile flow fields respectively. Three
or four reach a limit cycle, with relative phases oscillating in a
way reminiscent of cilia.

We note that self T-dual swimmers and pumps, such as the three-sphere
swimmer with equal arm lengths, are forbidden by symmetry from phase
locking. This agrees with the result of Kim and Powers \cite{kim_hydrodynamic_2004}
who showed that two rigid helices with parallel axes, driven by the
same torque, do not synchronise.

Phase synchronisation in the form of metachronal waves, which are
thought to be stabilised by hydrodynamic interactions, is well established
in cilliary dynamics. It would be interesting to see whether any similar
behaviour is apparent in systems of biological swimmers. Moreover,
phase-locking will affect the behviour of fabricated micropumps, and
questions remain as to whether synchronisation can enhance flow.

\section{Acknowledgements}

We thank G.P. Alexander, J. Dunkel, S.A. Edwards, C.M. Pooley, and
I. Zaid for helpful discussions.

\section{Bibliography}

\bibliographystyle{apsrev}
\bibliography{library}

\end{document}